\documentclass[11pt, letterpaper]{article}
\usepackage[utf8]{inputenc}
\usepackage[margin=1.5cm]{geometry}
\usepackage{titlesec}
\usepackage{tabu}
\usepackage{enumitem}
\usepackage{amssymb}
\usepackage{xcolor}
\usepackage{booktabs}
\usepackage{multirow}
\usepackage{graphicx}
\usepackage{textpos}
\usepackage[acronym]{glossaries}
\newlist{selectlist}{itemize}{2}
\setlist[selectlist]{label=$\square$,leftmargin=*,noitemsep,topsep=0pt}
\usepackage{tikz}
\usetikzlibrary{positioning, arrows.meta}

\newacronym{gui}{GUI}{Graphical User Interface}
\newacronym{crs}{CRS}{Coordinate Reference Systems}
\newacronym{cir}{CIR}{Color Infra-Red}
\newacronym{png}{PNG}{Portable Network Graphics}
\newacronym{dtm}{DTM}{Digital Terrain Model}
\newacronym{dsm}{DSM}{Digital Surface Model}
\newacronym{dem}{DEM}{Digital Elevation Model}
\newacronym{gis}{GIS}{Geographic Informaiton System}
\newacronym{uav}{UAV}{Unmanned Aerial Vehicle}
\newacronym{ca}{CA}{Cellular Automata}
\newacronym{ndvi}{NDVI}{Normalized Difference Vegetation Index}

\usepackage{lmodern}

\usepackage{hyperref}
\hypersetup{
    colorlinks,
    linkcolor={black!50!black},
    citecolor={black!50!black},
    urlcolor={black!80!black}
}

\titleformat{\section}[block]{\hspace{1em}\bfseries}{\thesection.}{0.5em}{}
\titleformat{\subsection}[block]{\hspace{1em}}{\thesubsection}{0.5em}{}

\begin{document}

\graphicspath{{./figs/}}

\begin{textblock*}{190mm}(-1cm,-1.5cm)
  \noindent \footnotesize The peer-reviewed version of this paper is
  published in Software Impacts at
  \href{https://doi.org/10.1016/j.simpa.2024.100657}{\color{blue}https://doi.org/10.1016/j.simpa.2024.100657.}
  This version is typeset by the authors and differs only in pagination and typographical detail.
  \end{textblock*}

\noindent
\textbf{\textit{Raster Forge: Interactive Raster Manipulation Library and GUI for Python}}
\vskip0.5cm
\noindent
\textbf{\textit{Afonso Oliveira$^{1}$(afonso.oliveira@ulusofona.pt), Nuno Fachada$^{1,2}$, João P. Matos-Carvalho$^{1,2}$}}\\

\noindent
$^1$COPELABS, Lusófona University, Campo Grande 376, Lisbon, 1749-024, Lisbon, Portugal \\
\noindent
$^2$Center of Technology and Systems (UNINOVA-CTS) and Associated Lab of Intelligent Systems (LASI), 2829-516 Caparica, Portugal \\

\noindent
\textbf{Abstract}\\

Raster Forge is a Python library and graphical user interface for raster data manipulation and analysis. The tool is focused on remote sensing applications, particularly in wildfire management. It allows users to import, visualize, and process raster layers for tasks such as image compositing or topographical analysis. For wildfire management, it generates fuel maps using predefined models. Its impact extends from disaster management to hydrological modeling, agriculture, and environmental monitoring. Raster Forge can be a valuable asset for geoscientists and researchers who rely on raster data analysis, enhancing geospatial data processing and visualization across various disciplines.

\vskip0.5cm

\vskip0.5cm
\noindent
\textbf{Code Metadata}\\

\noindent
\begin{tabular}{|l|p{6.5cm}|p{9.5cm}|}
\hline
\textbf{Nr.} & \textbf{Code metadata description} & \textbf{Please fill in this column} \\
\hline
C1 & Current code version & v0.6.1 \\
\hline
C2 & Permanent link to code/repository used for this code version & \url{https://github.com/afe-oliveira/raster-forge} \\
\hline
C3  & Permanent link to Reproducible Capsule & N/A\\
\hline
C4 & Legal Code License & MIT License \\
\hline
C5 & Code versioning system used & Git \\
\hline
C6 & Software code languages, tools, and services used & Python\\
\hline
C7 & Compilation requirements, operating environments \& dependencies & Python 3.8+, PySide6, NumPy, Rasterio, Spyndex, OpenCV Python, Matplotlib \\
\hline
C8 & If available Link to developer documentation/manual & \url{https://afe-oliveira.github.io/raster-forge/} \\
\hline
C9 & Support email for questions & a22308561@alunos.ulht.pt \\
\hline
\end{tabular}\\
\vskip0.5cm
\noindent

\section{Motivation and Significance}
\label{sec:motivation}

Raster grids are one of the most common data structures for image representation \cite{sonka2013image, rafael2018digital}. Constructed as a matrix of cells or \textit{pixels}, they are also often associated with the storage, processing and representation of spatial data. Their importance lies in their ability to allow direct matrix computations, thus enabling several important processing techniques, such as segmentation and classification \cite{li2022real,chen2024fire,wang2024firevitnet}, resampling \cite{wang2024novel} and interpolation \cite{sanabria2013spatial}. Multispectral images are particularly suited for representation within raster grids, as each cell can store intensity values across different wavelengths. Moreover, this structure is also very significant for its ability to provide an approximate representation of spatial phenomena, especially when coupled with appropriate geographic information \cite{goodchild2009geographic}.

Python is an extremely versatile language with thousands upon thousands of software for a wide variety of applications. Among these are some very influential tools that allow the manipulation of raster data, from powerful libraries such as OSGeo GDAL \cite{gdal}, Rasterio \cite{gillies_2019} and OpenCV \cite{opencv_library}, to fully fledged \glspl{gis} like the GRASS GIS \cite{neteler2012grass}, ArcGIS \cite{arcgis2011} and QGIS \cite{QGIS_software}.

While these tools are undoubtedly powerful and feature-rich, they can present certain barriers to entry. For instance, some may find them difficult to install or use, especially those new to \gls{gis} or lacking technical expertise. Even seemingly straightforward libraries like Rasterio or OpenCV can present hurdles, as their functionalities often require some degree of research before successful implementation. In addition, the costs of licensing proprietary software such as ArcGIS can be prohibitive for individuals or organizations with limited budgets.

\begin{figure}
    \centering
    \begin{tikzpicture}[>=Stealth, node distance=1cm]

        \node[] (central) {\includegraphics[width=5cm]{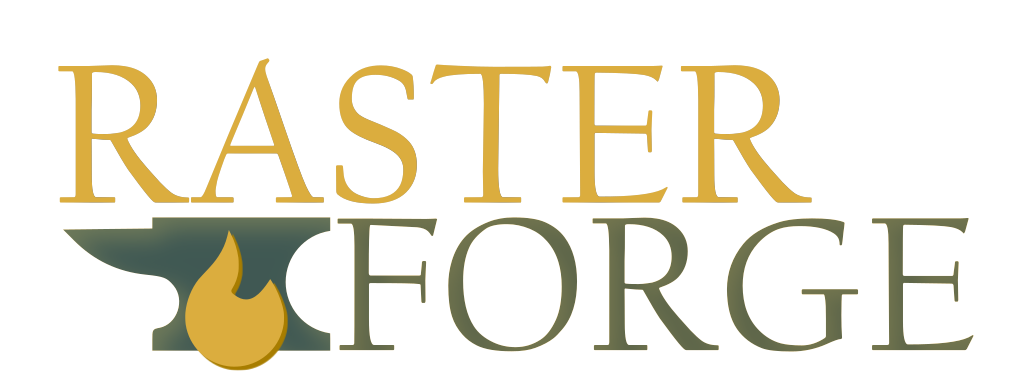}};

        \node[above right=of central] (numpy) {\includegraphics[width=3cm]{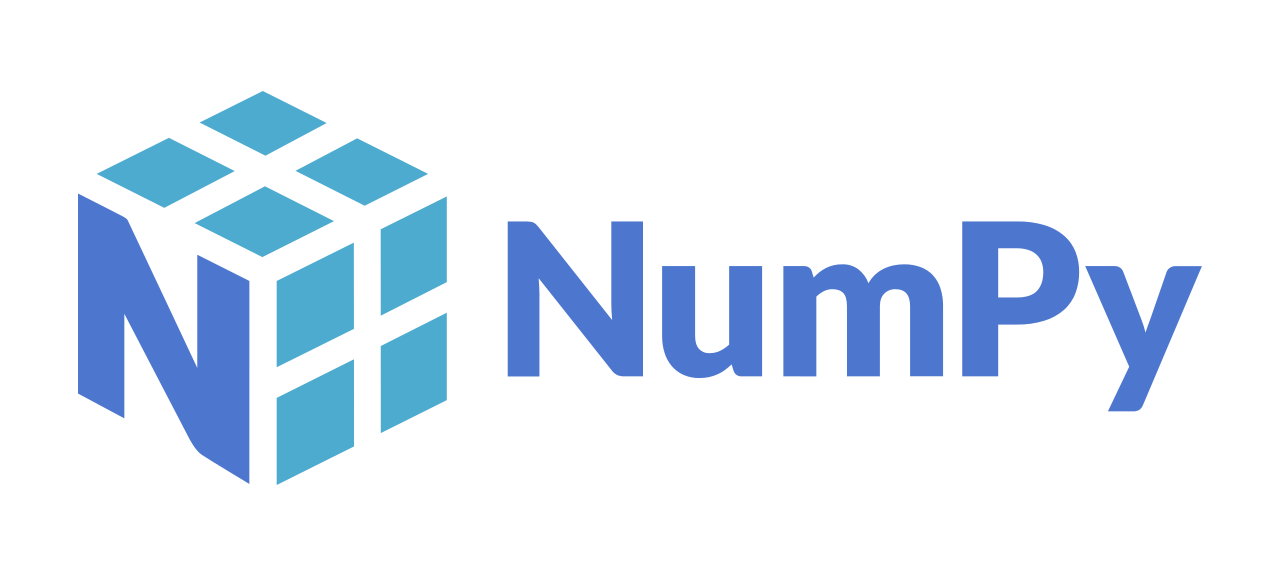}};
        \node[above left=of central] (rasterio) {\includegraphics[width=1.5cm]{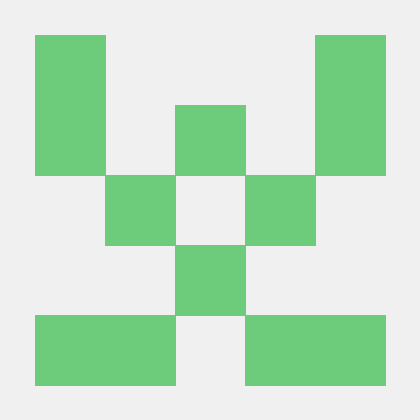}};
        \node[below right=of central] (spyndex) {\includegraphics[width=3cm]{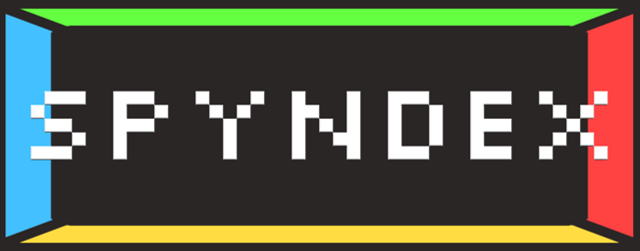}};
        \node[below left=of central] (opencv) {\includegraphics[width=1.75cm]{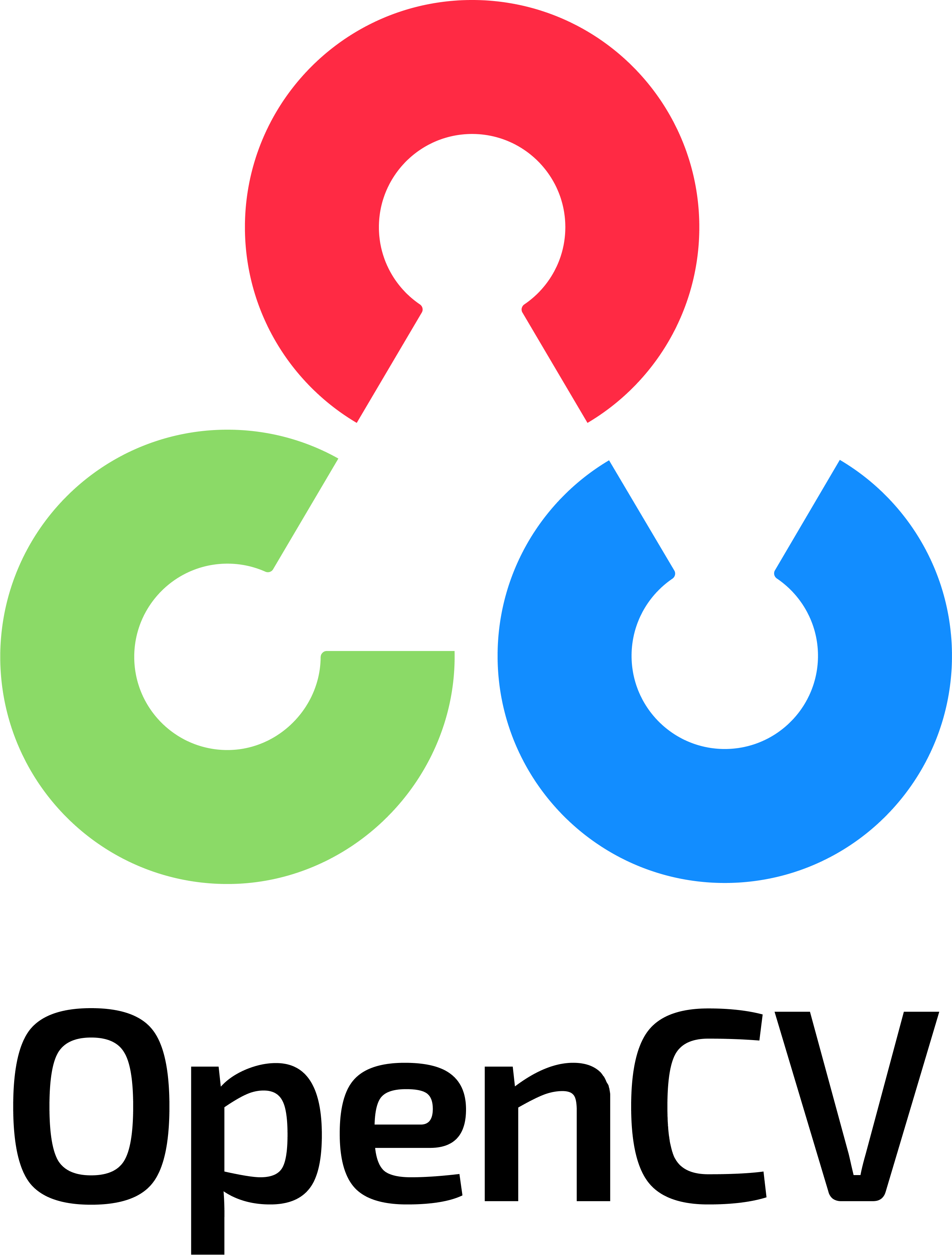}};
        \node[above=of central] (pyside6) {\includegraphics[width=2.5cm]{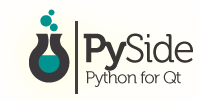}};

        \draw[->] (numpy) -- (central);
        \draw[->] (rasterio) -- (central);
        \draw[->] (spyndex) -- (central);
        \draw[->] (opencv) -- (central);
        \draw[->] (pyside6) -- (central);

        \node[below=0.05cm of rasterio] {Rasterio};
        \node[above=0.05cm of pyside6] {PySide6};
    \end{tikzpicture}
    \caption{Major libraries used to build Raster Forge. Adapted from \cite{rasterio-git,pyside-logo,numpypng,spyndex-logo,opencv-media-kit}.}
    \label{fig:flowchart}
\end{figure}
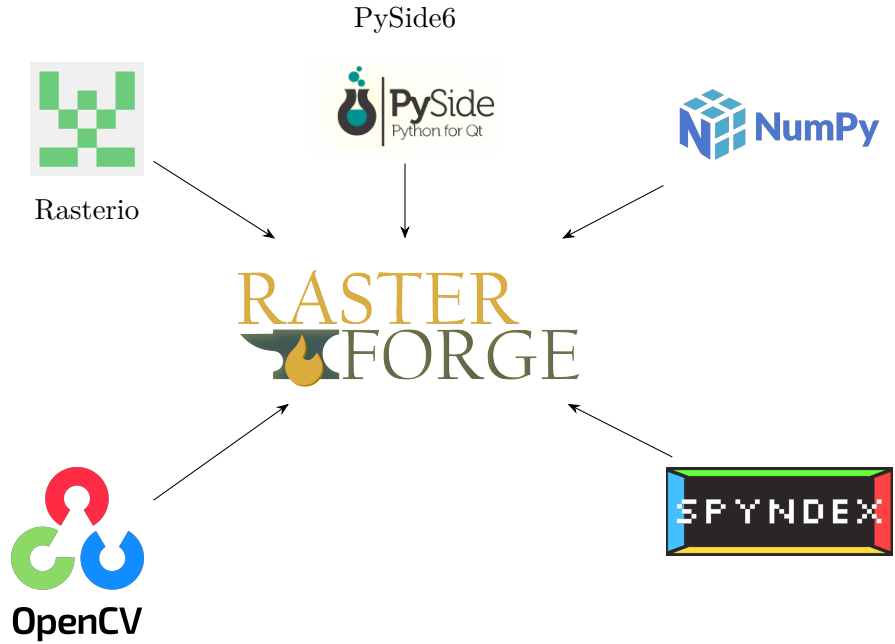

In this paper we present Raster Forge, an easy to use library and \gls{gui}, which aims to fill this gap for a beginner-friendly raster manipulation tool for spatial analysis. The library is powerful, direct and thoroughly documented, making its usage very straightforward. Additionally, its intuitive \gls{gis} eliminates the need for extensive research beforehand, allowing users to employ it quickly and effortlessly.

Raster Forge places great emphasis on the geospatial capabilities inherent in the raster structure. Although primarily designed to aid in wildfire management, care has been taken to ensure its versatility for more general applications.

\section{Software Description}
\label{sec:description}

Raster Forge is divided into two parts: the library and the \gls{gui}. The library of implemented functions is detailed in Subsection~\ref{sub:library}, while the \gls{gui} application is presented in Subsection~\ref{sub:gui}.

\subsection{Library}
\label{sub:library}

The library component is comprised of two container classes, \texttt{Layer} and \texttt{Raster}, which are used to represent a single grid of information and a collection of such grids, respectively. These classes are described with additional detail in Subsection~\ref{sub:containers}. The processing functions are primarily designed to work with \texttt{Layer} objects, but they also support NumPy arrays (\texttt{NDArray}) \cite{harris2020array} as inputs and outputs.

The library's raster manipulation capabilities are divided into six categories, as outlined in Table~\ref{tab:raster-manipulation-processes}. These categories include the generation of image composites (Subsection~\ref{sub:composites}), multispectral indices (Subsection~\ref{sub:indices}), topographical features (Subsection~\ref{sub:topography}), distance fields (Subsection~\ref{sub:distance-field}), height maps (Subsection~\ref{sub:height}), and fuel maps (Subsection~\ref{sub:fuel}).

\begin{table}
\begin{center}
\caption{Library raster manipulation processes. \texttt{NDArray} is an array type provided by NumPy \cite{harris2020array} and represents a multidimensional, homogeneous array of fixed-size items.} \label{tab:raster-manipulation-processes}
\footnotesize
\begin{tabular}{p{2cm} p{4cm} p{4cm}}
    \toprule
    \multirow{2}{*}{\textbf{Task}} & \multicolumn{2}{l}{\textbf{Inputs}} \\
    \cmidrule(rl){2-3}
    & \textbf{Name} & \textbf{Data Type} \\
    \midrule
    \multirow{4}{*}{Composites} & Layers & \textit{List[\texttt{NDArray}, \texttt{Layer}]} \\
    & Alpha & \textit{\texttt{NDArray}, \texttt{Layer}} \\
    & Gamma & \textit{List[float], Tuple[float]} \\
    & As Array & \textit{bool} \\
    \midrule
    \multirow{6}{*}{Indices$^a$} & Index ID & \textit{String} \\
    & Index Parameters & \textit{Dictionary} \\
    & Alpha & \textit{\texttt{NDArray}, \texttt{Layer}} \\
    & Thresholds & \textit{List[float], Tuple[float]} \\
    & Binarize & \textit{bool} \\
    & As Array & \textit{bool} \\
    \midrule
    \multirow{4}{*}{Slope} & DEM & \textit{\texttt{NDArray}, \texttt{Layer}} \\
    & Units & \textit{degrees, radians} \\
    & Alpha & \textit{\texttt{NDArray}, \texttt{Layer}} \\
    & As Array & \textit{bool} \\
    \cmidrule(rl){2-3}
    \multirow{4}{*}{Aspect} & DEM & \textit{\texttt{NDArray}, \texttt{Layer}} \\
    & Units & \textit{degrees, radians} \\
    & Alpha & \textit{\texttt{NDArray}, \texttt{Layer}} \\
    & As Array & \textit{bool} \\
    \midrule
    \multirow{6}{*}{Distance Field} & \texttt{Layer} & \textit{\texttt{NDArray}, \texttt{Layer}} \\
    & Alpha & \textit{\texttt{NDArray}, \texttt{Layer}} \\
    & Thresholds & \textit{List[float], Tuple[float]} \\
    & Invert & \textit{bool} \\
    & Mask Size & \textit{3, 5} \\
    & As Array & \textit{bool} \\
    \midrule
    \multirow{4}{*}{Height} & DTM & \textit{\texttt{NDArray}, \texttt{Layer}} \\
    & DSM & \textit{\texttt{NDArray}, \texttt{Layer}} \\
    & Distance & \textit{\texttt{NDArray}, \texttt{Layer}} \\
    & Alpha & \textit{\texttt{NDArray}, \texttt{Layer}} \\
    & As Array & \textit{bool} \\
    \midrule
    \multirow{9}{*}{Fuel Map} & Vegetation Coverage & \textit{\texttt{NDArray}, \texttt{Layer}} \\
    & Canopy Height & \textit{\texttt{NDArray}, \texttt{Layer}} \\
    & Distance & \textit{\texttt{NDArray}, \texttt{Layer}} \\
    & Water Features Mask & \textit{\texttt{NDArray}, \texttt{Layer}} \\
    & Artificial Structures Mask & \textit{\texttt{NDArray}, \texttt{Layer}} \\
    & Fuel Models & \textit{List[int], Tuple[int]} \\
    & Tree Height & \textit{float} \\
    & Alpha & \textit{\texttt{NDArray}, \texttt{Layer}} \\
    & As Array & \textit{bool} \\
    \bottomrule
    & & \\
    \multicolumn{3}{l}{\parbox{10cm}{$^a$ The multispectral indices architecture is built around the Spyndex Python library \cite{montero2023standardized}.}}
\end{tabular}
\end{center}
\end{table}

\subsubsection{Containers}
\label{sub:containers}

The \texttt{Layer} class holds the data and metadata of a single raster map, including geographic information and direct statistics, as shown in Table~\ref{tab:layer-properties}. The architecture of this object ensures that all relevant information is easily accessible through its properties. A data import function has also been developed to facilitate the extraction of a single band as a \texttt{Layer} from a raster file (see Table~\ref{tab:layer-methods}).

\begin{table}[h]
\begin{center}
\caption{Layer object properties.\label{tab:layer-properties}}
\footnotesize
\begin{tabular}{p{3cm} p{5cm} }
    \toprule
    \textbf{Name} & \textbf{Data Type} \\
    \midrule
    Array & \textit{\texttt{NDArray}} \\
    Bounds & \textit{Dictionary} \\
    CRS & \textit{String} \\
    Driver & \textit{String} \\
    No Data & \textit{int, float} \\
    Transform & \textit{(float, float, float, float, float, float)} \\
    Units & \textit{String} \\
    \midrule
    Resolution & \textit{int} \\
    Width & \textit{int} \\
    Height & \textit{int} \\
    Count & \textit{int} \\
    \midrule
    Mean & \textit{float} \\
    Median & \textit{float} \\
    Min & \textit{int, float} \\
    Max & \textit{int, float} \\
    Standard Deviation & \textit{float} \\
    \bottomrule
    \end{tabular}
\end{center}
\end{table}

\begin{table}[h]
\begin{center}
\caption{\texttt{Layer} object methods.\label{tab:layer-methods}}
\footnotesize
\begin{tabular}{p{2.5cm} p{2cm} p{2cm} p{7.5cm}}
    \multirow{2}{*}{\textbf{Task}} & \multicolumn{2}{l}{\textbf{Inputs}} & \multirow{2}{*}{\textbf{Description}} \\
    \cmidrule(rl){2-3}
    & Name & Data Type & \\
    \midrule
    \multirow{3}{*}{Import Layer} & File Path & \textit{String} & \multirow{3}{*}{\parbox{7.5cm}{Imports one band from a specified file as a layer. Applies scale to dataset if specified.}} \\
    & Band ID & \textit{int} & \\
    & Scale & \textit{int} & \\
    \bottomrule
    \end{tabular}
\end{center}
\end{table}

The \texttt{Raster} class acts as a container for a group of
\texttt{Layer}s and is associated with an uniform pixel size---or scale---that applies to all constituent layers for uniformity. The properties of this class are described in Table~\ref{tab:raster-properties}, and its main responsibilities include the bulk import of raster data and the manipulation of the \texttt{Layer} collection, as highlighted in Table~\ref{tab:raster-methods}.

\begin{table}[h]
\begin{center}
\caption{Raster object properties.\label{tab:raster-properties}}
\footnotesize
\begin{tabular}{p{2.5cm} p{5cm}}
    \toprule
    \textbf{Name} & \textbf{Data Type} \\
    \midrule
    Layers & \textit{Dictionary[String: \texttt{Layer}]} \\
    Scale & \textit{int} \\
    \bottomrule
    \end{tabular}
\end{center}
\end{table}

\begin{table}[h]
\begin{center}
\caption{Raster object methods.\label{tab:raster-methods}}
\footnotesize
\begin{tabular}{p{2.5cm} p{2cm} p{3cm} p{6.5cm}}
    \toprule
    \multirow{2}{*}{\textbf{Task}} & \multicolumn{2}{l}{\textbf{Inputs}} & \multirow{2}{*}{\textbf{Description}} \\
    \cmidrule(rl){2-3}
    & Name & Data Type & \\
    \midrule
    \multirow{2}{*}{Import Layers} & File Path & \textit{String} & \multirow{2}{*}{\parbox{6.5cm}{Imports a series of bands as defined on the \textit{Config} dictionary.}} \\
    & Config & \textit{Dictionary} & \\
    \midrule
    \multirow{2}{*}{Add Layer} & Name & \textit{String} & \multirow{2}{*}{\parbox{6.5cm}{Adds a \textit{\texttt{Layer}} to the \textit{Layers} Dictionary.}} \\
    & Layer & \textit{\texttt{Layer}, \texttt{NDArray}} & \\
    Remove Layer & Name & \textit{String} & \parbox{6.5cm}{Remove a \textit{\texttt{Layer}} from the \textit{Layers} Dictionary.} \\
    \multirow{2}{*}{Edit Layer} & Old Name & \textit{String} & \multirow{2}{*}{\parbox{6.5cm}{Changes the name identifier of a given \textit{\texttt{Layer}} in the \textit{Layers} dictionary.}} \\
    & New Name & \textit{String} & \\
    \bottomrule
    \end{tabular}
\end{center}
\end{table}

\subsubsection{Composites}
\label{sub:composites}

This feature allows the creation of true and false colour composites. It includes gamma correction functionality, allowing the user to select the gamma value to be applied to each individual layer through a power law transformation, $I_{\text{out}} = I_{\text{in}}^\gamma$, where \(I_{\text{out}}\) represents the output intensity or colour value, \(I_{\text{in}}\) represents the input intensity or colour value, and \(\gamma\) represents the gamma value applied for adjustment.

\subsubsection{Multispectral Indices}
\label{sub:indices}

Users can choose from a broad range of indices provided by the Spyndex \cite{montero2023standardized} package and compute them directly using defined Layers. It also allows the application of a threshold to the index output, which facilitates the generation of binary masks.

\subsubsection{Topographical Features}
\label{sub:topography}

A user is able compute both the slope and the aspect (orientation of slopes) of a region defined by \gls{dem} data. This implementation utilizes the direct slope calculation formula given by equation \ref{eq:slope} and the aspect calculation given by equation \ref{eq:aspect}. In these equations, the variable $\text{array}$ represents a set of values, usually a \gls{dem} or a grid that represents the surface. The variables $x$ and $y$ represent the spatial dimensions or coordinates within the array.

\begin{equation} \label{eq:slope}
\arctan \left( \sqrt{ \left( \frac{\partial \text{{array}}}{\partial x} \right)^2 + \left( \frac{\partial \text{{array}}}{\partial y} \right)^2 } \right)
\end{equation}

\begin{equation} \label{eq:aspect}
\arctan \left( -\frac{\partial \text{{array}}}{\partial x}, \frac{\partial \text{{array}}}{\partial y} \right)
\end{equation}

\subsubsection{Distance Field}
\label{sub:distance-field}

The distance field is computed from the given dataset, and each pixel of the image is labelled with the distance to the nearest obstacle pixel (non-zero pixel). The data can also be pre-binarized through a provided threshold before processing.

\subsubsection{Height Map}
\label{sub:height}

Generates a map that displays the vertical distance between the Earth's surface, as recorded in a \gls{dtm}, and the highest recorded point, as represented in a \gls{dsm}. This spatial data provides a three-dimensional perspective to flat vegetation coverage maps.

\subsubsection{Fuel Map}
\label{sub:fuel}

Combining vegetation cover, distance fields, canopy height, water features and man-made structures data, and a standard tree height benchmark, this functionality can produce a detailed fuel map for the geographic region of interest. The process integrates three distinct fuel models, including solely vegetative, standalone trees, and mixed vegetation, resulting in a depiction of fuel distribution across the landscape.

\subsection{Graphical User Interface (GUI)}
\label{sub:gui}

\begin{figure}
\begin{center}
\footnotesize
\begin{tabular}{c}
    \includegraphics[width=0.75\textwidth]{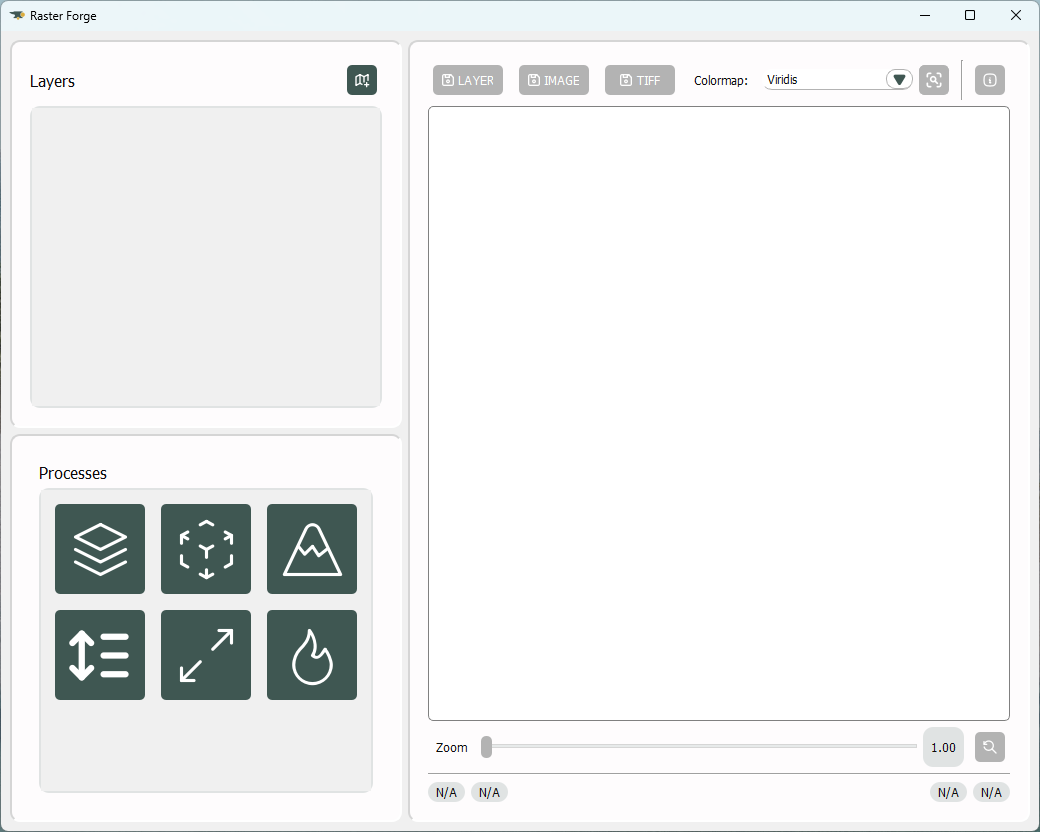} \\
    \end{tabular}
\caption{The \gls{gui} is arranged into three distinct panels: the layers panel located at the top-left, the processes panel positioned at the bottom-left, and the viewer panel situated on the right. \label{fig:gui}}
\end{center}
\end{figure}

The \gls{gui} has been developed using PySide6 \cite{pyside6} and comprises three distinct panels, as shown in Figure~\ref{fig:gui}. One panel is dedicated to importing data and managing Layers, another panel facilitates configuring and triggering processes, while the third panel serves as a data viewer, allowing users to visualize the layers being worked on. In this context, a \textit{layer} refers to an array or group of arrays, which are treated as cohesive data groups.

\begin{figure}
\begin{center}
\footnotesize
\begin{tabular}{cc}
    \includegraphics[width=.4\textwidth]{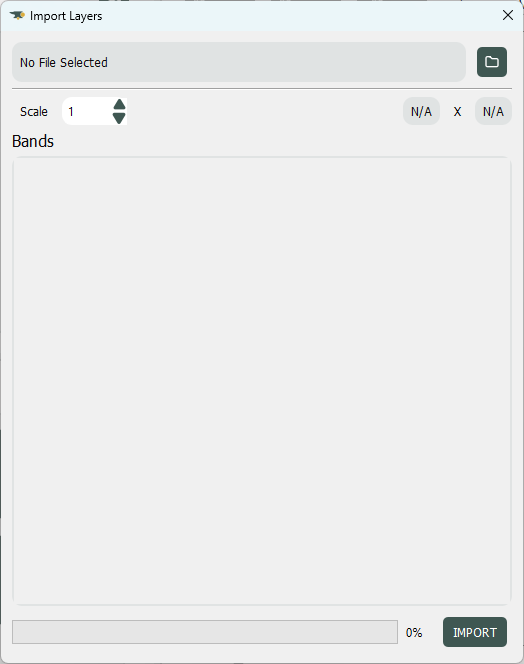} &
    \includegraphics[width=.4\textwidth]{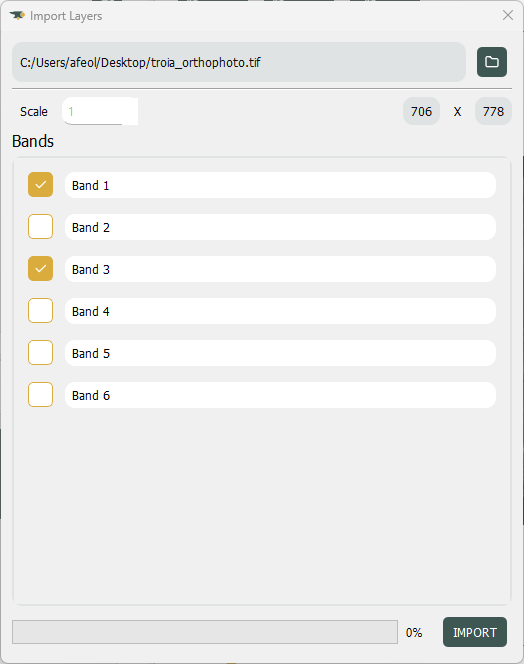} \\
     (a) & (b) \\
    \end{tabular}
\caption{\texttt{Layer} import panel showcasing its initial state (a) and after selection of a file (b). The predicted values for width and height dynamically adjust based on the chosen scale. \label{fig:import}}
\end{center}
\end{figure}

To simplify importing data, a dedicated window has been designed to enable users to select a file, as depicted in Figure~\ref{fig:import}. Within this window, users can preview the file's composition, including the number of bands it contains. Users can select bands to import and specify the desired scale. Additionally, the window displays the predicted width and height of the resulting array, aiding users in assessing the computational resources required for any further processing.

\begin{figure}
\begin{center}
\footnotesize
\begin{tabular}{c}
    \includegraphics[width=.755\textwidth]{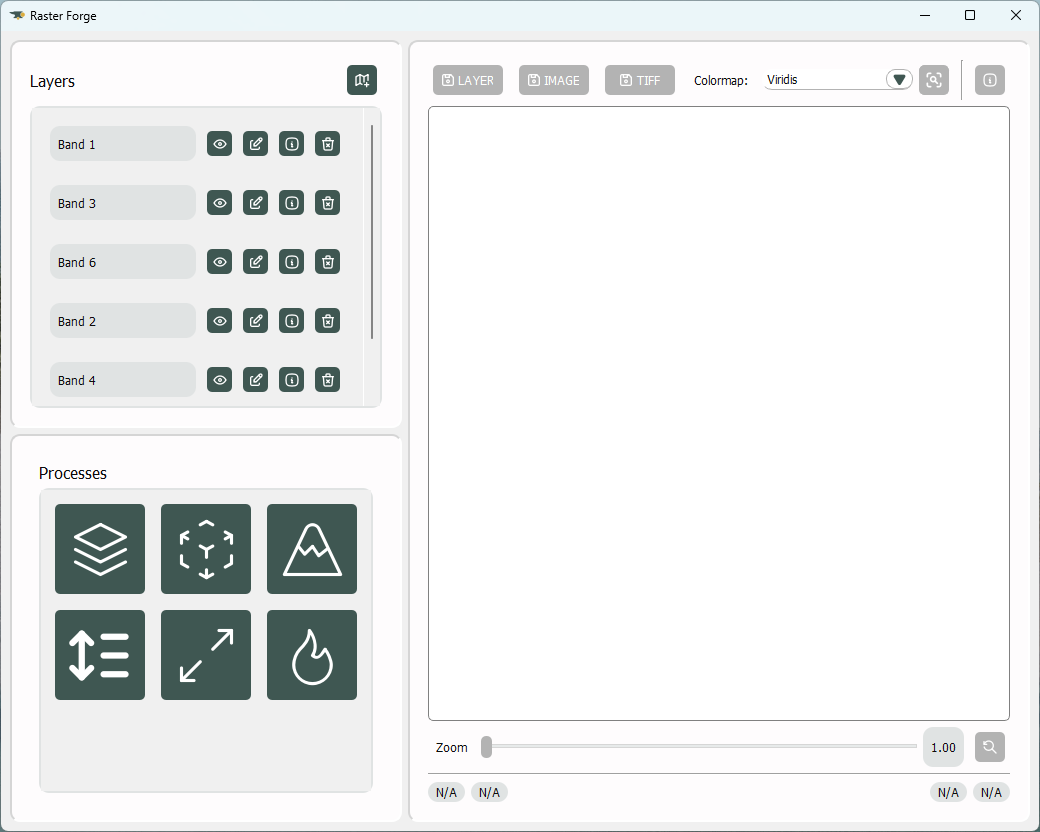} \\
    \end{tabular}
\caption{Upon importing layers, the graphical user interface (\gls{gui}) offers access to layer functionalities. For each layer, four functionalities are available, namely: view layer, edit layer's name, access layer information, and delete layer.\label{fig:gui-layers}}
\end{center}
\end{figure}

\begin{figure}
\begin{center}
\footnotesize
\begin{tabular}{ccc}
    \includegraphics[width=.25\textwidth]{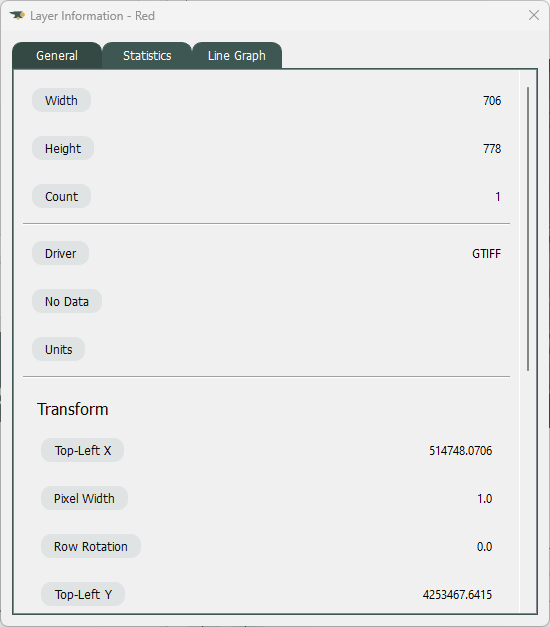} &
    \includegraphics[width=.25\textwidth]{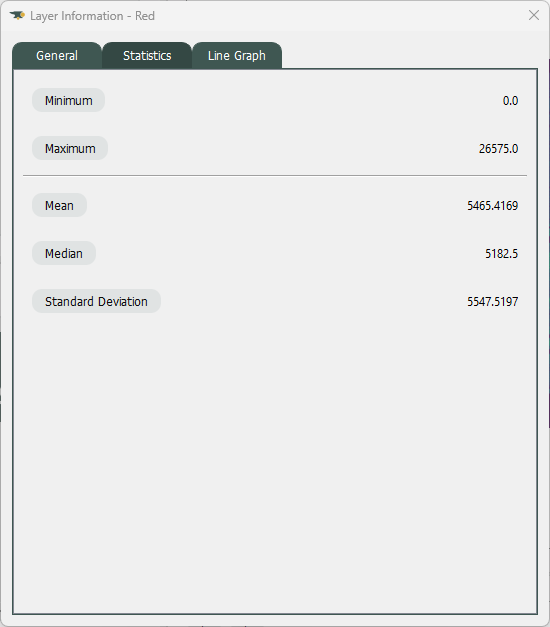} &
    \includegraphics[width=.25\textwidth]{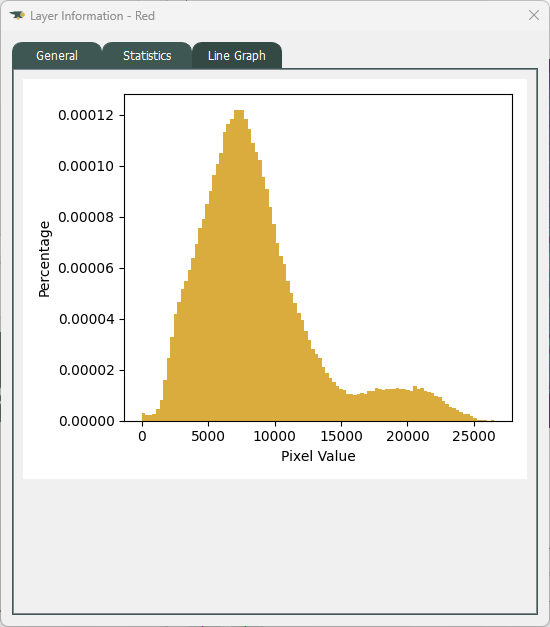} \\
     (a) & (b) & (c)\\
    \end{tabular}
\caption{\texttt{Layer} information panel composed of three tabs: metadata, statistical insights, and value histogram.\label{fig:gui-layers-info}}
\end{center}
\end{figure}

After the layers are imported, they are displayed on the \texttt{Layer} list, as depicted in Figure~\ref{fig:gui-layers}. Each layer offers four distinct functionalities: viewing the layer, editing the layer identifier, accessing layer information, and deleting the layer. \texttt{Layer} information is presented in an information window (see Figure~\ref{fig:gui-layers-info}) consisting of three tabs: metadata, statistical data, and a value histogram.

\begin{figure}
\begin{center}
\footnotesize
\begin{tabular}{cc}
    \includegraphics[width=.4\textwidth]{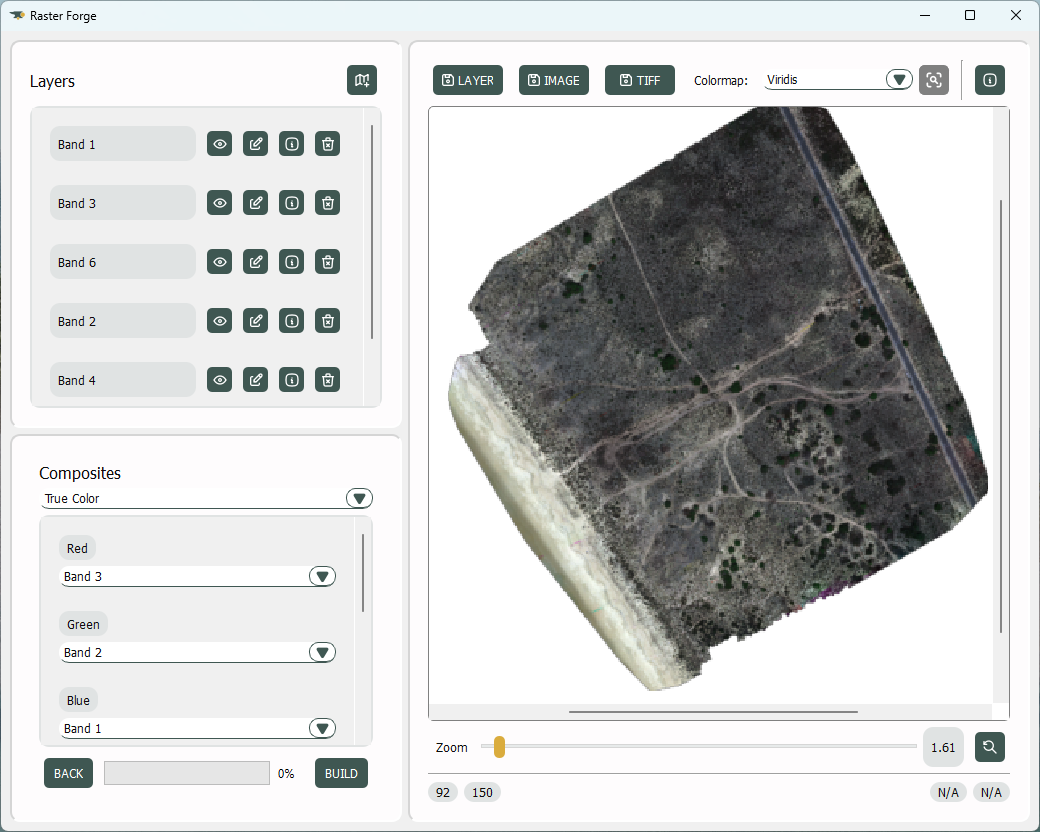} &
    \includegraphics[width=.4\textwidth]{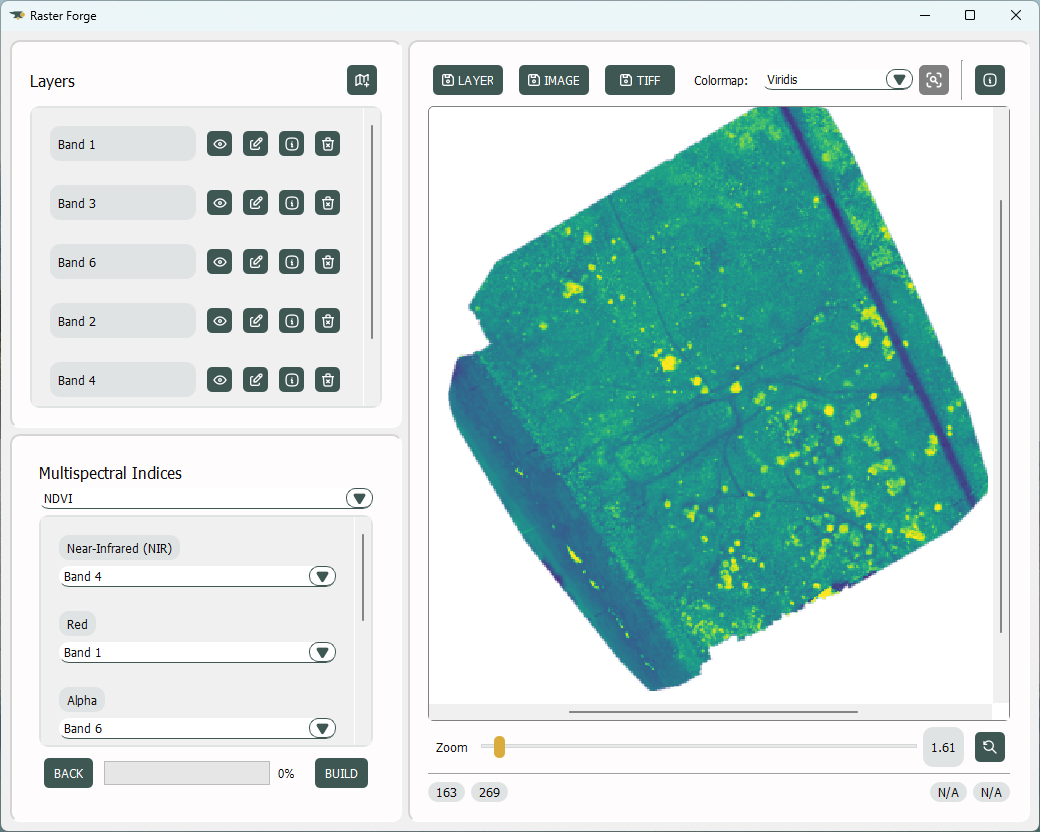} \\
     (a) & (b) \\
    \end{tabular}
\caption{Creating a (a) true color composite, and a (b) multispectral index map. The adaptive panel inputs are located within the processes panel, alongside the functionalities accessible within the view panel once an image is being viewed. These include zoom options, colormap selection, and various saving actions\label{fig:gui-use}}
\end{center}
\end{figure}

Once the layers are loaded, they can be used in various processes. All processing functions in the library are accessible through the \gls{gui}. When selecting a process, an adaptive panel displays all necessary inputs. The process can be initiated once all inputs are fulfilled, as shown in Figure~\ref{fig:gui-use}. Once activated, the data is displayed on the viewer and can be saved in various formats: as a layer for reuse in other processes, as an image with the applied colormap, or as a raw TIFF file containing all geographical information of the area. The information window for the currently viewed data is available directly on the viewer panel, even if it has not been saved as a layer yet.

\section{Impact}
\label{sec:impact}

Raster Forge has had impact in a few projects already. The software has been used to compute terrain feature maps from \gls{uav} imagery \cite{oliveira2024data}. This permitted to easily showcase results including multispectral index and vegetation coverage maps, as well as water and man-made structures masks. An early iteration of this software was also used to compute fuel type maps---along with the components necessary to build those maps, such as canopy height, vegetation coverage and soil bareness---that were integral in processing \gls{uav} data for testing/validation of a wildfire simulation model \cite{oliveira2023decision}. Finally, in a similar way to the previous solution, the library component of this software is also currently being used in the development of a web wildfire simulation application under the DAIS Project \cite{dais}.

Expected Raster Forge short-term future impact centers around providing faster processing of large-scale aerial data for several ongoing studies. The software is being utilized to aid the construction of topographical and land-cover datasets for quickly producing terrain feature maps, such as elevation, canopy height, and vegetation density. These datasets will then be employed to train a diffusion model with the aim of generating synthetic topographical and land-cover data for wildfire model testing. In addition, Raster Forge is also being used to generate fuel maps and their sub-components (distance fields, vegetation density, man-made structures, and water feature masks, etc.) to test and validate in-development land cover and fuel type classification methods. These fuel maps are also being employed as auxiliary data to support the conceptualization and preliminary testing of a novel time and space continuous \gls{ca} wildfire model. The impact on the described studies is already being felt, as the ease-of-use and capabilities of Raster Forge are significantly expediting the data harmonization stage, thus substantially reducing the expected development times.

From the emphasis placed on the geographical aspect of each raster, Raster Forge potential impact will likely be felt in the field of remote sensing, particularly where it intersects with wildfire management. However, we believe that Raster Forge has the capability to streamline raster data processing across many other fields of study in which raster data manipulation is a fundamental concept, including medical imaging \cite{latif2019medical,jardim2023image}, architecture \cite{nasim2023veernet, tu2023integrating}, and planetary \cite{li2023machine,huang2021commentary} and interplanetary \cite{tao2023star,zhang2021infrared} remote sensing, broadening its potential impact.

\section{Conclusions}
\label{sec:conclusions}

In conclusion, Raster Forge provides a solution for manipulating raster data, specifically designed for remote sensing applications such as wildfire management. The Raster Forge library offers a range of functions for raster processing, including composite generation, multispectral index computation, topographical feature extraction, and fuel map creation. Its included \gls{gui} is user-friendly, streamlines the use of the functionalities of the library and simplifies data import, layer management, process configuration, and result visualization. This enhances accessibility for users with varying levels of expertise. Raster Forge has potential applications in diverse fields such as disaster management, hydrological modeling, agriculture, and environmental monitoring, positioning itself as a valuable tool for geospatial data analysis and visualization tasks.

\section*{Funding}
This work was financed by the Portuguese Agency FCT (Funda\c{c}\~ao para a Ci\^encia e Tecnologia), in the framework of projects UIDB/00066/2020, UIDB/04111/2020, CEECINST/00002/2021/CP2788/CT0001 and CEECINST/00147/2018/CP1498/CT0015.

\vskip0.3cm
\noindent

\bibliographystyle{unsrt}

\end{document}